\documentclass{aa}
\usepackage{graphicx}

\begin{document}

\title{\object{LS~5039}: a runaway microquasar ejected from the galactic plane}
%\titlerunning{\object{LS~5039}: a microquasar ejected from the galactic plane}

\author{M. Rib\'o\inst{1}
\and J.~M. Paredes\inst{1}
\and G.~E. Romero\inst{2}
\and P. Benaglia\inst{2}
\and J. Mart\'{\i}\inst{3}
\and O. Fors\inst{1}
\and J. Garc\'{\i}a-S\'anchez\inst{1}}

\offprints{M. Rib\'o, \email{mribo@am.ub.es}}

\institute{Departament d'Astronomia i Meteorologia, Universitat de
Barcelona, Av. Diagonal 647, 08028 Barcelona, Spain
\and Instituto Argentino de Radioastronom\'{\i}a, C.C.5, (1894) Villa Elisa, Buenos Aires, Argentina
%\and Max-Planck-Institut f\"ur Kernphysik, Postfach 10 39 80, 69029 Heidelberg, Germany
\and Departamento de F\'{\i}sica, Escuela Polit\'ecnica Superior, Universidad de Ja\'en, Virgen de la Cabeza 2, 23071 Ja\'en, Spain
}

\date{Received 10 December 2001 / Accepted 10 January 2002}

\abstract{
We have compiled optical and radio astrometric data of the microquasar
\object{LS~5039} and derived its proper motion. This, together with the
distance and radial velocity of the system, allows us to state that this source
is escaping from its own regional standard of rest, with a total systemic
velocity of about $150$~km~s$^{-1}$ and a component perpendicular to the
galactic plane larger than 100~km~s$^{-1}$. This is probably the result of an
acceleration obtained during the supernova event that created the compact
object in this binary system. We have computed the trajectory of
\object{LS~5039} in the past, and searched for OB associations and supernova
remnants in its path. In particular, we have studied the possible association
between \object{LS~5039} and the supernova remnant \object{G016.8$-$01.1},
which, despite our efforts, remains dubious. We have also discovered and
studied an \ion{H}{i} cavity in the ISM, which could have been created by the
stellar wind of \object{LS~5039} or by the progenitor of the compact object in
the system. Finally, in the symmetric supernova explosion scenario, we estimate
that at least $17\,M_{\sun}$ were lost in order to produce the high
eccentricity observed. Such a mass loss could also explain the observed runaway
velocity of the microquasar.
\keywords{
stars: individual: \object{LS~5039}, \object{RX~J1826.2$-$1450} --
X-rays: binaries -- 
supernovae: individual: \object{G016.8$-$01.1} -- 
radio continuum: stars --
radio continuum: ISM -- 
ISM: supernova remnants
} 
}

\maketitle

\section{Introduction} \label{introduction}

Microquasars are stellar-mass black holes or neutron stars that mimic, on
smaller scales, many of the phenomena seen in AGN and quasars. These objects
have been found in X-ray binary systems, where a compact object accretes matter
from a companion star. Radio emitting X-ray binaries with relativistic radio
jets, like \object{SS~433}, \object{GRS~1915+105}, \object{GRO~J1655$-$40} or
\object{Cygnus~X-3}, are good examples of microquasars (see Mirabel \&
Rodr\'{\i}guez \cite{mirabel99} for a detailed review). With the recent
addition of \object{LS~5039} (Paredes et~al. \cite{paredes00}),
\object{Cygnus~X-1} (Stirling et~al. \cite{stirling01}) and
\object{XTE~J1550$-$564} (Hannikainen et~al. \cite{hannikainen01}) to the
microquasar group, the current number of this kind of sources is 14, among
$\sim280$ known X-ray binaries (Liu et~al. \cite{liu00}; Liu et~al.
\cite{liu01}) of which $\sim50$ display radio emission. Recent studies of
microquasars can be found in Castro-Tirado et~al. (\cite{castro01}).

Attention to the star \object{LS~5039} was first called by Motch et~al.
(\cite{motch97}), who proposed it as a High Mass X-ray Binary (HMXB) candidate
associated with the X-ray source \object{RX~J1826.2$-$1450}. The object is
located at an estimated distance of $\sim3.1$~kpc and close to the galactic
plane ($l=16.88\degr$, $b=-1.29\degr$). Soon after, non-thermal and moderately
variable radio emission was reported by Mart\'{\i} et~al. (\cite{marti98})
using the Very Large Array (VLA). The evidence of its microquasar nature was
provided by Paredes et~al. (\cite{paredes00}) when radio jets were discovered
with the Very Long Baseline Array (VLBA) at milliarcsecond (mas) angular
scales. These authors also pointed out the possible connection of
\object{LS~5039} with \object{3EG~J1824$-$1514}, i.e., one of the unidentified
EGRET sources of high energy $\gamma$-rays. X-ray observations of
\object{RX~J1826.2$-$1450} by Rib\'o et~al. (\cite{ribo99}) did not reveal
pulsations and were consistent with a significantly hard X-ray spectrum up to
30~keV, with a strong Gaussian iron line at 6.6~keV.

The mass donor in \object{LS~5039} was originally classified as an O7V((f))
star by Motch et~al. (\cite{motch97}). This classification has been recently
improved thanks to optical and near infrared spectroscopic observations by
Clark et~al. (\cite{clark01}), which indicate an O6.5V((f)) star.

%------------------------------------------------------------------------------
\begin{table*}
\begin{center}
\caption[]{Compilation of optical and radio positions, with associated errors, of \object{LS~5039}.}
\label{positions}
\begin{tabular}{lllclcll}
\hline \noalign{\smallskip}
Wavelength & Epoch & $\alpha$~(ICRS) & $\sigma_{\alpha\cos\delta}$ & $\delta$~(ICRS)               & $\sigma_\delta$ & Catalog & Reference\\
Domain &      & (h, m, s)       & (mas)                       & ($\degr$, $\arcmin$, $\arcsec$) & (mas)           & name    &\\
\noalign{\smallskip} \hline \noalign{\smallskip}
Optical & 1905.45  & 18 26 15.0194 & 255 & -14 50 53.300 & 247 & AC~2000.2 & Urban, private communication\\
        & 1907.43  & 18 26 15.0177 & 255 & -14 50 53.075 & 247 & AC~2000.2 & Urban, private communication\\
        & 1951.577 & 18 26 15.034  & 250 & -14 50 53.59  & 250 & USNO-A2.0 & Monet et~al. \cite{monet99}\\
        & 1979.484 & 18 26 15.0557 & ~64 & -14 50 54.075 & ~47 & TAC~2.0   & Zacharias \& Zacharias \cite{zacharias99a}\\
        & 1986.653 & 18 26 15.054  & 300 & -14 50 54.29  & 300 & GSC~1.2   & Morrison et~al. \cite{morrison01}\\
        & 1991.75  & 18 26 15.0427 & 149 & -14 50 54.229 & 120 & Tycho-2   & H{\o}g et~al. \cite{hog00}\\
        & 2000.289 & 18 26 15.0563 & ~13 & -14 50 54.277 & ~13 & UCAC1     & Zacharias et~al. \cite{zacharias00}\\
\noalign{\smallskip} \hline \noalign{\smallskip}
Radio   & 1998.24  & 18 26 15.056  & ~10 & -14 50 54.24  & ~10 & VLA obs.  & Mart\'{\i} et~al. \cite{marti98}\\
        & 2000.42  & 18 26 15.0566 & ~~4 & -14 50 54.261 & ~~6 & VLBA obs. & Rib\'o et~al. \cite{ribo02}\\
\noalign{\smallskip} \hline
\end{tabular}
\end{center}
\end{table*}
%------------------------------------------------------------------------------

Our present knowledge of the system orbit is based only on the radial velocity
measurements in the optical by McSwain et~al. (\cite{mcswain01}). Their most
remarkable findings consist of a short orbital period of $P=4.117$~days and a
significant eccentricity of $e=0.41$. They were also able to determine the
system radial velocity and a mass function of $f(m)=0.00103\,M_{\sun}$. The
fact that the available optical photometry does not show an ellipsoidal
modulation could indicate that the orbital inclination is very low. Therefore,
the possibility of having a black hole in the system cannot be ruled out in
spite of the small mass function observed.

The formation of the compact object in an X-ray binary necessarily requires a
supernova explosion that does not disrupt the system. This explosive event is
expected to considerably change the kinematic properties of the binary star.
The change may be rather extreme (i.e., kick velocities approaching $\sim10^3$
km~s$^{-1}$) for highly asymmetric supernovae, as it has been proposed for the
microquasar \object{Circinus~X-1} (Tauris et~al. \cite{tauris99}). This
mechanism may also be responsible for ejecting X-ray binaries into the halo of
the Galaxy. The fast moving X-ray nova \object{XTE~J1118+480} is a possible
example (Mirabel et~al. \cite{mirabel01}), although it could have been ejected
from a globular cluster in the past. On the other hand, the microquasars
\object{GRO~J1655$-$40} and \object{Cygnus~X-1}, also display runaway
velocities (Shahbaz et~al. \cite{shahbaz99}; Kaper et~al. \cite{kaper99}). With
independence of their origin in the galactic plane or in the halo, the
existence of runaway microquasars is a new and important issue that deserves an
in-depth study.

In this paper we focus on the kinematic properties of \object{LS~5039} and its
surroundings. In Sect.~\ref{ppm} we estimate the proper motions of the system,
using optical and radio data. In Sect.~\ref{ejection} we discuss the distance
and space velocity of \object{LS~5039}, whereas in Sect.~\ref{trajectory} we
analyze the trajectory of the binary system in the past. In Sect.~\ref{snr} we
study the possible association of this microquasar with a supernova remnant,
and in Sect.~\ref{surroundings} we analyze their  \ion{H}{i} surroundings.
Finally, we make a global discussion in Sect.~\ref{discussion} and summarize
our main conclusions in Sect.~\ref{conclusions}.

\section{Positions and proper motions} \label{ppm}

\subsection{Optical positions} \label{optical}

\object{LS~5039} ($V\simeq11.2, B\simeq12.2)$ is not included in most accurate
astrometric catalogs, like Hipparcos and Tycho-1 (which are complete up to
$V\sim9.0$ and $V\sim10.0$, respectively). Although it appears in Tycho-2, its
position has a large uncertainty when compared to the average error within its
magnitude range (H{\o}g et~al. \cite{hog00}). As a result, the catalogued
proper motions lack of the required precision. Moreover, other astrometric
catalogs which were not included in Tycho-2 estimate of the proper motions,
have been released since then. In view of all these facts, we decided to carry
out a thorough study of all available catalogs. First, we did a search using
the query forms within the VizieR database (Ochsenbein et~al.
\cite{ochsenbein00}) at the Centre de Donn\'ees astronomiques de Strasbourg
(CDS), and afterwards we looked for new information on several public web
sites. The results of this search are listed in Table~\ref{positions}, with all
coordinates in ICRS, and commented below.

The oldest astrometric information for \object{LS~5039} is dated around 1905
and comes from the Astrographic Catalog 2000 (AC~2000, Urban et~al.
\cite{urban98}). However, a re-reduction of this catalog was performed
recently, using an improved reference catalog that allowed a better handling of
the systematic errors (S.~E. Urban, private communication). This new catalog
will soon be released as AC~2000.2. Since changes were most significant in the
faint stars of the southern AC~2000 zones, the position of \object{LS~5039}
suffered a noticeable shift. Actually, two positions corresponding to different
epochs are given in AC~2000.2.

The following available astrometry, dated around 1950, comes from the scanning
and reduction of the Palomar Observatory Sky Survey plates, carried out by
Monet et~al. (\cite{monet99}), which is part of the USNO-A2.0 catalog.

We inspected the Second Cape Photographic Catalog (CPC2, Zacharias et~al.
\cite{zacharias99b}), a Southern Hemisphere astrometric catalog, containing
observations between 1962 and 1973. Unfortunately, \object{LS~5039} is not
listed in it, because the limiting magnitude of this catalog was $V\simeq10.5$.

The Twin Astrographic Catalog version 2 (TAC~2.0) provides positional
information around 1980. Apart from the internal error of
$\sigma_{\alpha\cos\delta}=62$~mas and $\sigma_\delta=45$~mas associated to the
\object{LS~5039} position, an external error of 15~mas has been taken into
account following an estimate carried out by Zacharias \& Zacharias
(\cite{zacharias99a}).

The last photographic catalog used is the Guide Star Catalog version 1.2
(GSC~1.2, Morrison et~al. \cite{morrison01}). This is a re-reduction of the GSC
(Lasker at~al. \cite{lasker90}), after removing plate-based systematic
distorsions that are a function of magnitude and radial distance from the plate
center. In addition, by using the {\it Starlink} library {\tt SLALIB}, we have
transformed the catalogued position from FK5 to ICRS.

As mentioned above, \object{LS~5039} appears as well in the Tycho-2 catalog
(H{\o}g et~al. \cite{hog00}). The average internal error in position for a
$V\simeq11.2$ star is around 50~mas, while the errors in $\alpha\cos\delta$ and
$\delta$ for \object{LS~5039} are 93 and 120~mas, respectively. This fact tells
us that the position estimate is not as good as one would expect. On the other
hand, the scatter-based errors from the 4 positions present in the Tycho-2 data
are 149 and 60~mas, respectively. Hence, we have considered
$\sigma_{\alpha\cos\delta}=149$~mas and $\sigma_\delta=120$~mas.

A second epoch of the GSC was performed during the nineties. However, the
positional errors in the current version of this catalog, GSC~2.2.01, are
intended as indicators for operational use only, and cannot be used for
scientific studies. Hence, we have considered that this catalog does not
provide accurate astrometric information to be used for estimating proper
motions.

Finally, we have used the first release of the US Naval Observatory CCD
Astrograph Catalog (UCAC1, Zacharias et~al. \cite{zacharias00}). As can be seen
in Table~\ref{positions}, this is by far the best optical astrometric catalog
available.

\subsection{Radio positions} \label{radio}

Independent astrometric information can be obtained from radio observations of
\object{LS~5039}. The first available astrometric information comes from the
NVSS by Condon et~al. (\cite{condon98}). However, the position error in this
catalog, around 1$\arcsec$, is too large to be of any use for our purposes.
Mart\'{\i} et~al. (\cite{marti98}) carried out VLA\footnote{The VLA and the
VLBA are operated by the National Radio Astronomy Observatory (NRAO). The NRAO
is a facility of the National Science Foundation operated under cooperative
agreement by Associated Universities, Inc.}-A configuration observations which
provided accurate (10~mas) astrometry for \object{LS~5039}. Finally,
phase-referencing VLBA$^1$ observations conducted by Rib\'o et~al.
(\cite{ribo02}) provide the last radio position, with the best accuracy among
all available data. Details of the astrometry for the last two cases are listed
in Table~\ref{positions}. The relatively large location error of few mas in the
VLBA data is due to confusion produced by the high level of Galactic electron
scattering in this region (for both the reference source and \object{LS~5039}).

We must state that no correction for parallax has been applied to the obtained
positions, since at an estimated distance of $\sim3$~kpc, this would translate
into less than half a mas, which is always much smaller than the available
uncertainties.

\subsection{Proper motions} \label{pm}

It is clear from the astrometric data that both, the optical and radio sources,
are almost in the same position of the sky. However, in order to show that this
is not a chance coincidence and that both emissions originate in the same
object, we have computed independent proper motions for the optical and radio
data.

Although the position uncertainties from the AC~2000.2 and USNO-A2.0 catalogs
are relatively large, the epoch span achieved justifies their inclusion in the
estimate of the proper motions. Hence, from the optical data, and taking into
account the astrometric uncertainties, we obtain the following results:
$\mu_{\alpha\cos\delta}=4.7\pm1.3$~mas~yr$^{-1}$,
$\mu_{\delta}=-11.0\pm0.8$~mas~yr$^{-1}$, where the errors come directly from
the least squares fit. The two accurate radio positions give proper motions of:
$\mu_{\alpha\cos\delta}=4.0\pm4.9$~mas~yr$^{-1}$,
$\mu_{\delta}=-9.6\pm5.3$~mas~yr$^{-1}$. Although this last result has a large
error because only two points are available, we can say that both, the optical
and radio sources, have very similar proper motions. Therefore, based only on
astrometric data, we are able to confirm that both, the optical and the radio
emission, originate in the same object. Hence, we will use all the data listed
in Table~\ref{positions} to compute accurate proper motions for
\object{LS~5039}, which happen to be:
$\mu_{\alpha\cos\delta}=4.7\pm1.1$~mas~yr$^{-1}$,
$\mu_{\delta}=-10.6\pm1.0$~mas~yr$^{-1}$. All these results are summarized in
Table~\ref{pmt}. We can now transform the proper motions into galactic
coordinates and obtain $\mu_{l}=-7.2\pm1.0$~mas~yr$^{-1}$ and
$\mu_{b}=-9.1\pm1.0$~mas~yr$^{-1}$. It is clear from these results that there
is a noticeable motion perpendicular to the galactic plane and moving away from
it.

%------------------------------------------------------------------------------
\begin{table}
\begin{center}
\caption[]{Proper motions estimates for \object{LS~5039}.}
\label{pmt}
\begin{tabular}{lcc}
\hline \noalign{\smallskip}
Data set      & $\mu_{\alpha\cos\delta}$ & $\mu_\delta$\\
              & (mas~yr$^{-1}$)          & (mas~yr$^{-1}$)\\
\noalign{\smallskip} \hline \noalign{\smallskip}
Optical       & $4.7\pm1.3$              & $-11.0\pm0.8$\\
Radio         & $4.0\pm4.9$              & ~\,$-9.6\pm5.3$\\
Optical+Radio & $4.7\pm1.1$              & $-10.6\pm1.0$\\
\noalign{\smallskip} \hline
\end{tabular}
\end{center}
\end{table}
%------------------------------------------------------------------------------

According to our least squares fits, and defining $t={\rm yr}-2000.0$, the
predicted ICRS values for $\alpha$ and $\delta$ near $t=0$ are:
\begin{equation}
{\alpha=18^{\rm h}~26^{\rm m}~15.0565^{\rm s}~+~\left[4.7\,t/\cos\delta\over{\rm mas}\right]}
\label{alpha}
\end{equation}
\begin{equation}
{\delta=-14\degr~50\arcmin~54.260\arcsec~-~\left[10.6\,t\over{\rm mas}\right]}
\label{delta}
\end{equation}
being $\sqrt{9+1.2\,t^2/\cos^2\delta}$ and $\sqrt{9+t^2}$ the errors in mas in
$\alpha$ and $\delta$, respectively. Comparing the fits with the positions in
Table~\ref{positions} and the proper motions in Table~\ref{pmt} we can say
that, approximately, the offsets are determined by the VLBA data point because
of the small error in position, while the proper motions are obtained by the
optical points due to the huge time span.

%------------------------------------------------------------------------------
\begin{figure*}
\resizebox{\hsize}{!}{\includegraphics{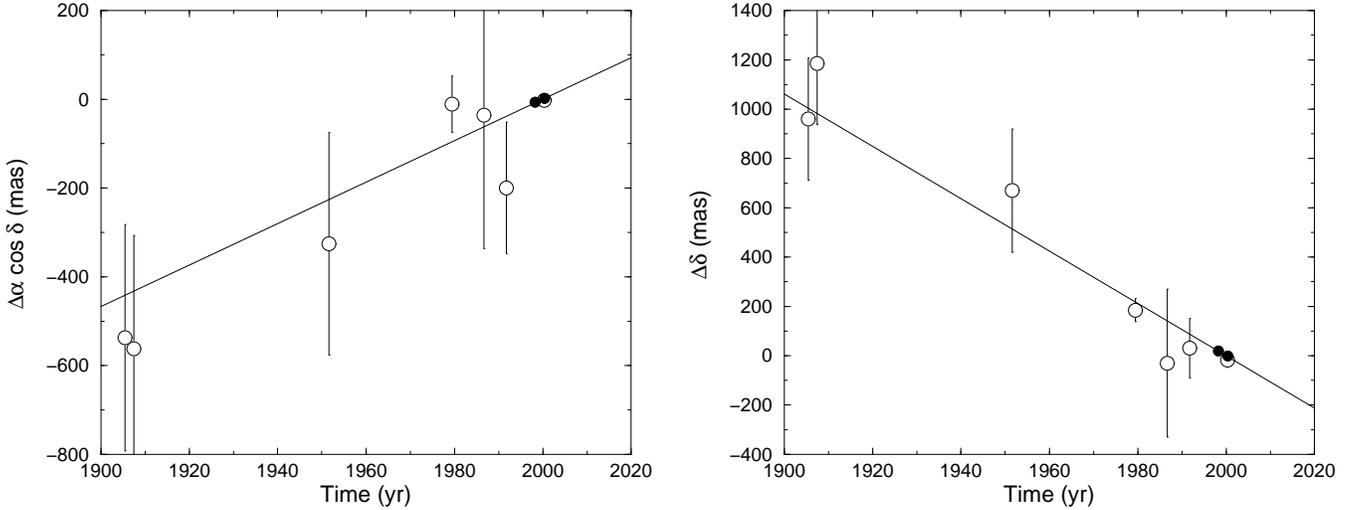}}
\caption{Offsets in Right Ascension (left) and Declination (right), relatives to the fitted ones for year 2000.0, versus time for all the positions listed in Table~\ref{positions}. Open circles represent the optical positions, and filled circles the radio ones. The solid lines represent the least squares fits to the whole data sets.}
\label{pmf}
\end{figure*}
%------------------------------------------------------------------------------

In Fig.~\ref{pmf} we have plotted the offsets in $\alpha\cos\delta$ and
$\delta$ from the fitted values for year 2000.0 versus time, for the data
listed in Table~\ref{positions}, together with the respective fits to the data.
Notice that in both plots the UCAC1 and VLBA data are almost superimposed.

\section{Ejection from the galactic plane} \label{ejection}

Since the distance to \object{LS~5039} and its uncertainty are fundamental to
compute the space velocity of the system, we think it is worth to perform first
of all a short study on this issue.

\subsection{Distance and its uncertainty} \label{distance}

Motch et~al. (\cite{motch97}) proposed a distance of 3.1~kpc to
\object{LS~5039} based on the star color excess, and stated that this estimate
could have a large uncertainty. This value was computed assuming an O7V((f))
spectral type. However, recent observations by Clark et~al. (\cite{clark01})
show that the optical companion has a spectral type of O6.5V((f)). On the other
hand, old calibrations for the intrinsic color index and absolute magnitude
(Johnson \cite{johnson66}, Deutschman et~al. \cite{deutschman76}) were used by
Motch et~al. (\cite{motch97}). Hence, we have performed a new estimate of the
distance taking into account the new spectral type and more recent
calibrations.

The optical photometry available up to now, containing at least $B$ and $V$
magnitudes comes from Drilling (\cite{drilling91}), Lahulla \& Hilton
(\cite{lahulla92}) and Clark et~al. (\cite{clark01}), and is listed in
Table~\ref{distancet}. Using an intrinsic color index of $(B-V)_0=-0.30\pm0.02$
for an O6.5V star (Schaerer et~al. \cite{schaerer96}, Lejeune \& Schaerer
\cite{lejeune01}), we can compute the color excess $E(B-V)$ for all the
observations. Finally, using the relationship
$A_V=(3.30+0.28\,(B-V)_0+0.04\,E_{B-V})\,E(B-V)$ (Schmidt-Kaler
\cite{schmidt82}) and $M_V=-4.99\pm0.3$ for an O6.5V star (Vacca et~al.
\cite{vacca96}) we obtain the distance estimates listed in
Table~\ref{distancet}. The weighted mean of these values is $2.9\pm0.3$~kpc,
and will be the distance to \object{LS~5039} used hereafter. Finally, assuming
that the Sun is at 8.5~kpc from the galactic center, we obtain a galactocentric
distance of $5.8\pm0.3$~kpc for \object{LS~5039}.

%------------------------------------------------------------------------------
\begin{table}
\begin{center}
\caption[]{Distance estimates to \object{LS~5039} derived from optical photometry.}
\label{distancet}
\begin{tabular}{ccc}
\hline \noalign{\smallskip}
$V$                & $B-V$             & $d$~(kpc)\\
\noalign{\smallskip} \hline \noalign{\smallskip}
$11.23\pm0.01\,^a$ & $0.95\pm0.01\,^a$ & $2.68\pm0.06$ \\
$11.20\pm0.02\,^b$ & $0.95\pm0.02\,^b$ & $2.64\pm0.06$ \\
$11.33\pm0.02\,^c$ & $0.85\pm0.02\,^c$ & $3.26\pm0.07$ \\
$11.32\pm0.01\,^d$ & $0.85\pm0.02\,^d$ & $3.25\pm0.07$ \\
\noalign{\smallskip} \hline
\end{tabular}
\end{center}
$^a$ Drilling (\cite{drilling91})\\
$^b$ Lahulla \& Hilton (\cite{lahulla92})\\
$^c$ 1996 October observations by Clark et~al. (\cite{clark01})\\
$^d$ 2000 September observations by Clark et~al. (\cite{clark01})\\
\end{table}
%------------------------------------------------------------------------------

\subsection{Space velocity} \label{space}

If we assume a distance of $2.9\pm0.3$~kpc to \object{LS~5039}, as seen in the
previous section, and a systemic $V_{\rm r}=4.6\pm0.5$~km~s$^{-1}$ (McSwain
et~al. \cite{mcswain01}), the proper motions estimates translate into
($U=40\pm5$, $V=-82\pm16$, $W=-118\pm19$) km~s$^{-1}$ in the Local Standard of
Rest (LSR) defined by ($U_{\sun}=9$, $V_{\sun}=12$, $W_{\sun}=7$) km~s$^{-1}$.
This gives a total systemic velocity of $v_{\rm
sys}=(U^2+V^2+W^2)^{1/2}=149\pm18$~km~s$^{-1}$. Using a value of
215~km~s$^{-1}$ for the galactic rotation at 5.8~kpc from the galactic center
(Fich et~al. \cite{fich89}) we can transform the galactic plane space
velocities ($U$ and $V$) from the LSR into the \object{LS~5039} regional
standard of rest (RSR). Applying this transformation we find $U=51\pm6$ and
$V=-71\pm16$~km~s$^{-1}$, and $v_{\rm sys}=147\pm17$~km~s$^{-1}$. All these
results are summarized in Table~\ref{svt}.

%------------------------------------------------------------------------------
\begin{table}
\begin{center}
\caption[]{Space velocity estimates, in km~s$^{-1}$, for \object{LS~5039} related to the LSR and to its RSR.}
\label{svt}
\begin{tabular}{lcccc}
\hline \noalign{\smallskip}
Frame & $U$      & $V$        & $W$         & $v_{\rm sys}$\\
\noalign{\smallskip} \hline \noalign{\smallskip}
LSR   & $40\pm5$ & $-82\pm16$ & $-118\pm19$ & $149\pm18$\\
RSR   & $51\pm6$ & $-71\pm16$ & $-118\pm19$ & $147\pm17$\\
\noalign{\smallskip} \hline
\end{tabular}
\end{center}
\end{table}
%------------------------------------------------------------------------------

Taking into account that the cosmic dispersion for early-type stars is
$(\sigma_U, \sigma_V, \sigma_W)\simeq(7, 8, 4)$~km~s$^{-1}$ (Torra et~al.
\cite{torra00}) we can conclude that this source is escaping from its own RSR
and has a large velocity component perpendicular to the galactic plane. Since
\object{LS~5039} is a HMXB containing an early-type star, it seems very
unlikely to be a high speed halo object crossing the galactic plane. Hence, the
most plausible explanation for the observed velocity is that the binary system
obtained an acceleration during the supernova event that created the compact
object in this microquasar.

\section{The past trajectory of \object{LS~5039}} \label{trajectory}

An interesting thing to do once a position and space velocity estimates are
available, is to compute the trajectory of \object{LS~5039} in the past, and
then to look for OB associations and SN Remnants (SNRs) in its path, in order
to establish possible relationships. For this purpose, we computed the galactic
orbital motion of this system under the gravitational field of the Galaxy. The
galactic mass model by Dauphole \& Colin (\cite{dauphole95}) was adopted for
the integration, which was was performed using a Runge-Kutta fourth-order
integrator using time steps of 1000 years.

It is helpful to search for OB associations in the past trajectory of runaway
X-ray binaries, as done by Ankay et~al. (\cite{ankay01}) with
\object{HD~153919}/\object{4U~1700$-$37}, because if evidences are found that
the runaway system originated in an OB association, it is possible to estimate
the age of the binary system after the SN explosion. According to the computed
trajectory in the past for \object{LS~5039}, there are two OB associations in
or close to its path in the plane of the sky: \object{Sct~OB3} and
\object{Ser~OB2} (Melnik \& Efremov \cite{melnik97}). Unfortunately, the
distance to \object{Sct~OB3} is $\sim1.5$~kpc (Melnik \& Efremov
\cite{melnik97}), and the distance to \object{Ser~OB2} is $1.9\pm0.3$~kpc
(Forbes \cite{forbes00}). Hence, the two OB associations found in the path of
\object{LS~5039} are too close to us to be related to it.

Since we have not been able to fix a limit on the integration time in the past
thanks to a possible relationship with an OB association, we will use the
vertical distribution of early-type stars in the Galaxy for this purpose.
O-type stars are typically located at distances within $45\pm20$~pc from the
galactic midplane (Reed \cite{reed00}). According to the galactic latitude of
\object{LS~5039} and using a distance of $2.9\pm0.3$~kpc, its actual height is
$Z=-65\pm7$~pc. Since it is escaping from the galactic plane, we can compute
the trajectory backward in time up to when it had a height of $Z=65$~pc, which
is $\sim1.1$~Myr ago. Hence, the SN explosion probably took place in the last
$\sim1.1$~Myr.

Thus, for a possible association of \object{LS~5039} with a SNR, we will focus
our attention on the galactic trajectory of the binary system for the last
$\sim1$~Myr. In addition, the likelihood of detecting a SNR decreases with
time, so it is not expected that the radio emission of the SNR can be observed
much beyond the time interval considered above (e.g., Shklovskii
\cite{shklovskii68}).

A search in a catalog of galactic supernova remnants (Green \cite{green00})
reveals the presence of several SNRs near the path of \object{LS~5039} on the
plane of the sky, though the center of none of them is crossed by its
trajectory. Only three of them, namely \object{SNR G016.7+00.1}, \object{SNR
G16.8$-$01.1}, and \object{SNR G018.8+00.3}, are within $1\degr$ of the
\object{LS~5039} path. The distance to \object{SNR G016.7+00.1} was estimated
to be at $\sim14$~kpc by Reynoso \& Mangum (\cite{reynoso00}), making
impossible an association with \object{LS~5039}. Dubner et~al.
(\cite{dubner99}) concluded that \object{SNR G018.8+00.3} is located at
$1.9\pm0.5$~kpc, and has an age of $\sim16\,000$~yr. Although the distance is
not very different from the one to \object{LS~5039}, the age of this SNR is by
far too short to be associated with the microquasar. Hence, the only remaining
candidate is \object{SNR G16.8$-$01.1}, which deserves a detailed study.

%------------------------------------------------------------------------------
\begin{figure}
\resizebox{\hsize}{!}{\includegraphics{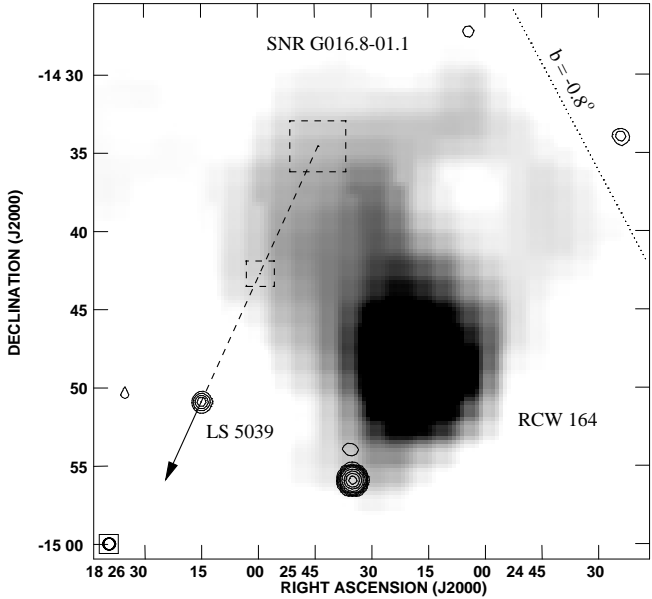}}
\caption{Wide field radio map of \object{LS~5039}, its nearby shell-like \object{SNR G016.8$-$01.1} and the \ion{H}{ii}-region \object{RCW~164} (which is the stronger source in the field). The grey scale emission is taken from the Parkes-MIT-NRAO tropical survey at the 6~cm wavelength (Tasker et~al. \cite{tasker94}). The overlaid contours correspond to the NVSS map of the region at the 20~cm wavelength (Condon et~al. \cite{condon98}). The arrow marks the proper motion sense (see text). The dashed line is the computed trajectory for the last $10^5$~yr, with the corresponding error boxes in position at that epoch and $5\,10^4$~yr ago. The dotted line represents positions with a galactic latitude of $-0.8\degr$ for reference purposes.}
\label{snrf}
\end{figure}
%------------------------------------------------------------------------------

\section{\object{SNR G016.8$-$01.1}} \label{snr}

A wide field radio map of the surrounding of \object{LS~5039} is shown in
Fig.~\ref{snrf}, with the microquasar (contours) and the nearby \object{SNR
G016.8$-$01.1} (grey scale), together with the position of the
\ion{H}{ii}-region \object{RCW~164} (which is the stronger source in the
field). The arrow in this figure marks the proper motion sense of
\object{LS~5039} as it would be seen from the LSR (i.e., correction for the
peculiar velocity of the Sun has been applied), while the dashed line
represents the trajectory in the past up to $10^5$~yr, with the corresponding
error boxes in position at that epoch and $5\,10^4$~yr ago. From this figure we
can see that \object{LS~5039} crosses the projection in the plane of the sky of
\object{SNR G016.8$-$01.1} between $\sim4\,10^4$ and $\sim1.3\,10^5$ years ago.

If one of the two stars forming a close binary system experiences a SN
explosion, it may form a SNR and an X-ray binary. Since conservation of the
linear momentum after the SN explosion is required, if the SNR and the X-ray
binary are related, we expect them to be aligned with the proper motion of the
latter, and any deviation from this behavior should be explained by the
projection in the plane of the sky of the initial space velocity of the binary
system prior to the SN event. This does not seem to be the case here, because
\object{SNR G016.8$-$01.1} and \object{LS~5039} are not well aligned with the
\object{LS~5039} proper motion. Moreover, if we assume that the center of the
shell-like remnant is located at the minimum of radio emission in
Fig.~\ref{snrf}, approximately located at $\alpha=18^{\rm h}~25^{\rm m}~03^{\rm
s}$, $\delta=-14\degr~37\arcmin~30\arcsec$ (or $l=16.94\degr$, $b=-0.93\degr$),
a peculiar velocity in the plane of the sky for the original system of
$\simeq70$~km~s$^{-1}$ is obtained. This velocity is much higher than the
typical $\sim10$ km~s$^{-1}$ found for early-type stars. In order to reduce
this velocity we can use in our favor the error in the proper motion angle, and
find that it turns out to be $\simeq50$~km~s$^{-1}$, which is still too high.
However, the center of the shell-like remnant could be in another position,
closer to the \object{LS~5039} trajectory, and the derived peculiar velocity
for the system prior to the SN explosion could fit in the typical values of
early-type stars. Hence, from the kinematical point of view, we are not able to
firmly discard a possible association between both objects. Therefore, we have
performed an in-depth study of the remnant.

\object{SNR G016.8$-$01.1} was discovered by Reich et~al. (\cite{reich86})
using a combination of survey data analysis and new multifrequency
observations. The extended radio source has a complex structure due to the
superposition of the \ion{H}{ii}-region \object{RCW~164} (Rodgers et~al.
\cite{rodgers60}), which is coincident with the peak of the continuum emission
(in black in Fig.~\ref{snrf}). The diffuse radiation observed around the peak
is strongly polarized (Rodgers et~al. \cite{rodgers60}), indicating a
synchrotron nature. Polarization decreases towards the position of the
foreground \ion{H}{ii}-region as a result of the thermal contribution. Thermal
and non-thermal components of the radiation cannot be clearly separated, but
the total flux from the SNR at 5~GHz seems to be $\sim1$~Jy (J.~A. Combi,
private communication). The angular diameter of the remnant is $\sim30\arcmin$;
its distance remains unknown.

\subsection{A lower limit of the distance} \label{lower}

\begin{figure}
\resizebox{\hsize}{!}{\includegraphics{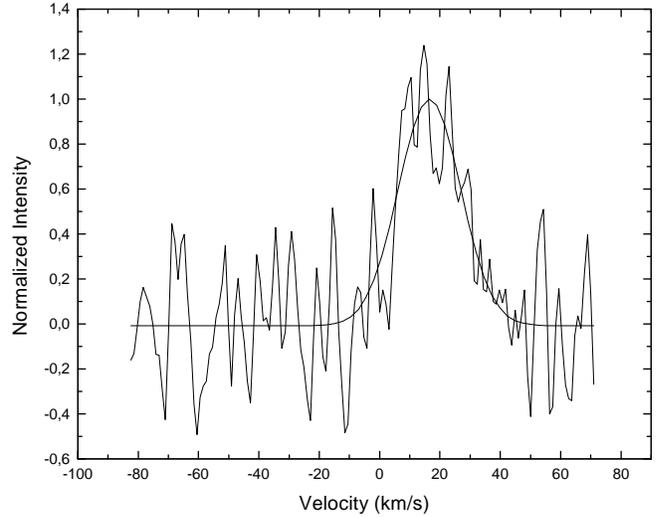}}
\caption{H166$\alpha$ line observations of the \ion{H}{ii}-region \object{RCW~164}. The solid line represents a Gaussian fit to the data, with its maximum being at $v=16.5\pm0.8$~km~s$^{-1}$.}
\label{line}
\end{figure}

In order to determine a lower limit of the distance to \object{SNR
G16.8$-$1.10}, we carried out H166$\alpha$ recombination line (1424.734~MHz)
observations of the foreground \ion{H}{ii}-region in November 17th 2000. We
used a 30-m radiotelescope at the Instituto Argentino de Radioastronom\'{\i}a
(IAR), Villa Elisa. The receiver is a helium-cooled HEMT amplifier with a
1008-channel autocorrelator at the back end. The HPBW at a wavelength of 21~cm
is $30\arcmin$ and the temperature of the system on the cold sky during the
observations was about 35~K. The H166$\alpha$ line was detected at a velocity
of $16.5\pm0.8$~km~s$^{-1}$ (see Fig.~\ref{line}) after 1.5 hours of
integration time, with a signal-to-noise ratio of $\sim4$. At a location of
$l\approx16.8\degr$, $b\approx-1.1\degr$, standard galactic rotation models
(Fich et~al. \cite{fich89}) indicate that the observed velocity corresponds to
a distance $d\sim1.8$~kpc. Consequently, \object{SNR G16.8$-$1.10} should be
farther than $\sim2$~kpc, as suggested by its relatively small angular size.

%------------------------------------------------------------------------------
%------------------------------------------------------------------------------

\subsection{Particle density estimates} \label{particle}

If we adopt for \object{SNR G016.8$-$01.1} the same distance to
\object{LS~5039}, we can make some estimates of the parameters that would
characterize the SNR in case of being associated with the X-ray binary. Using a
size of $30\arcmin$ and a distance of $2.9$~kpc, the inferred radius would be
$R=12.7$~pc and it would be, thus, still in the adiabatic expansion phase.
Using the standard Sedov (\cite{sedov59}) solutions, we can express the
particle density of the ambient medium as
\begin{equation}
n=4.44\,10^{-8}\,t^2\,E_{51}\,R_{1}^{-5}~{\rm cm}^{-3},
\end{equation}
where $t$ is the age of the remnant in years, $E_{51}$ is the original energy
release of the SN explosion in units of $10^{51}$~erg and $R_{1}$ is the
remnant radius in units of 10~pc. Since the binary system is bound after the SN
explosion, its total mass has to be larger than the mass of the SNR (van den
Heuvel \cite{vdheuvel78}), which, therefore, should be moving at least with the
same space velocity in the opposite direction. Hence, the maximum age of the
remnant if it were related to \object{LS~5039}, would be around $5\,10^4$~yr.
If we use this age and assume that $E_{51}=0.4$ (Spitzer \cite{spitzer98}), we
find an ambient density of $n\sim13.5$~cm$^{-3}$. Considering the uncertainty
in the distance, we can say that $8<n<23$~cm$^{-3}$. Although such number
densities would not be very unusual towards this direction of the inner part of
the Galaxy, we have made \ion{H}{i} observations in order to estimate the
typical column densities in the region of interest.

The \ion{H}{i} observations were performed during November 29th and December
1st 2000, with the already mentioned IAR telescope. The \ion{H}{i} line was
observed in a hybrid total-power mode with a sampling on a $0.25\degr$-lattice
around the position of \object{SNR G016.8$-$01.1}, covering a total of
$4\degr\times4\degr$. The velocity resolution obtained was 1.05~km~s$^{-1}$,
with a total coverage of $\pm450$~km~s$^{-1}$. The rms level in brightness
temperature was $\sim0.1$~K. Those channel maps in the velocity range from 10
to 50~km~s$^{-1}$ (corresponding to distances between 1.5 and 4.5~kpc) were
inspected looking for clouds or a local minimum that could be associated to the
SNR.

\begin{figure}
\resizebox{\hsize}{!}{\includegraphics{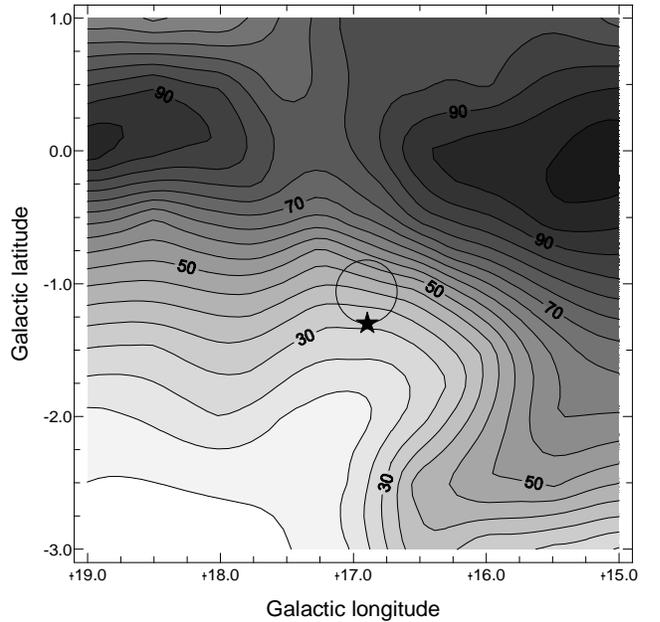}}
\caption{Neutral hydrogen column density distribution towards \object{SNR G016.8$-$01.1}, integrated over the velocity interval 26-34~km~s$^{-1}$. Contour levels are in units of 10$^{19}$~cm$^{-2}$. The circle marks the position of \object{SNR G016.8$-$01.1}, while the star represents \object{LS~5039}.}
\label{nhi}
\end{figure}

The maps show a strong gradient of brightness temperature towards the galactic
plane, but no clear evidence for a structure that could be associated with the
SNR. In Fig.~\ref{nhi}, we show the integrated column density map for the
velocity interval between 26-34~km~s$^{-1}$, where the contour labels are in
units of $10^{19}$~cm$^{-2}$. With the column densities observed in the region
where \object{SNR G016.8$-$01.1} is located ($\sim4\,10^{20}$~cm$^{-2}$), the
ambient density should be $\sim5$~cm$^{-3}$. This is nearly a factor of 3 lower
than the estimated density from the expected size of the SNR if it is at
2.9~kpc. Taking into account the uncertainty in the distance, the ambient
density determined from the remnant's size ($8<n<23$~cm$^{-3}$) would be
slightly larger than that estimated from the \ion{H}{i} observations. It could
be that there is more material under the form of other molecular species, or
that the original SN energy release could have been different from what we have
assumed, or, finally, that the distance to the SNR is not 2.9~kpc and it is not
related to \object{LS~5039}. The case, consequently, remains unsolved.

%------------------------------------------------------------------------------
%------------------------------------------------------------------------------

\section{The \ion{H}{i} surroundings of \object{LS~5039}} \label{surroundings}

\begin{figure*}
\resizebox{\hsize}{!}{\includegraphics{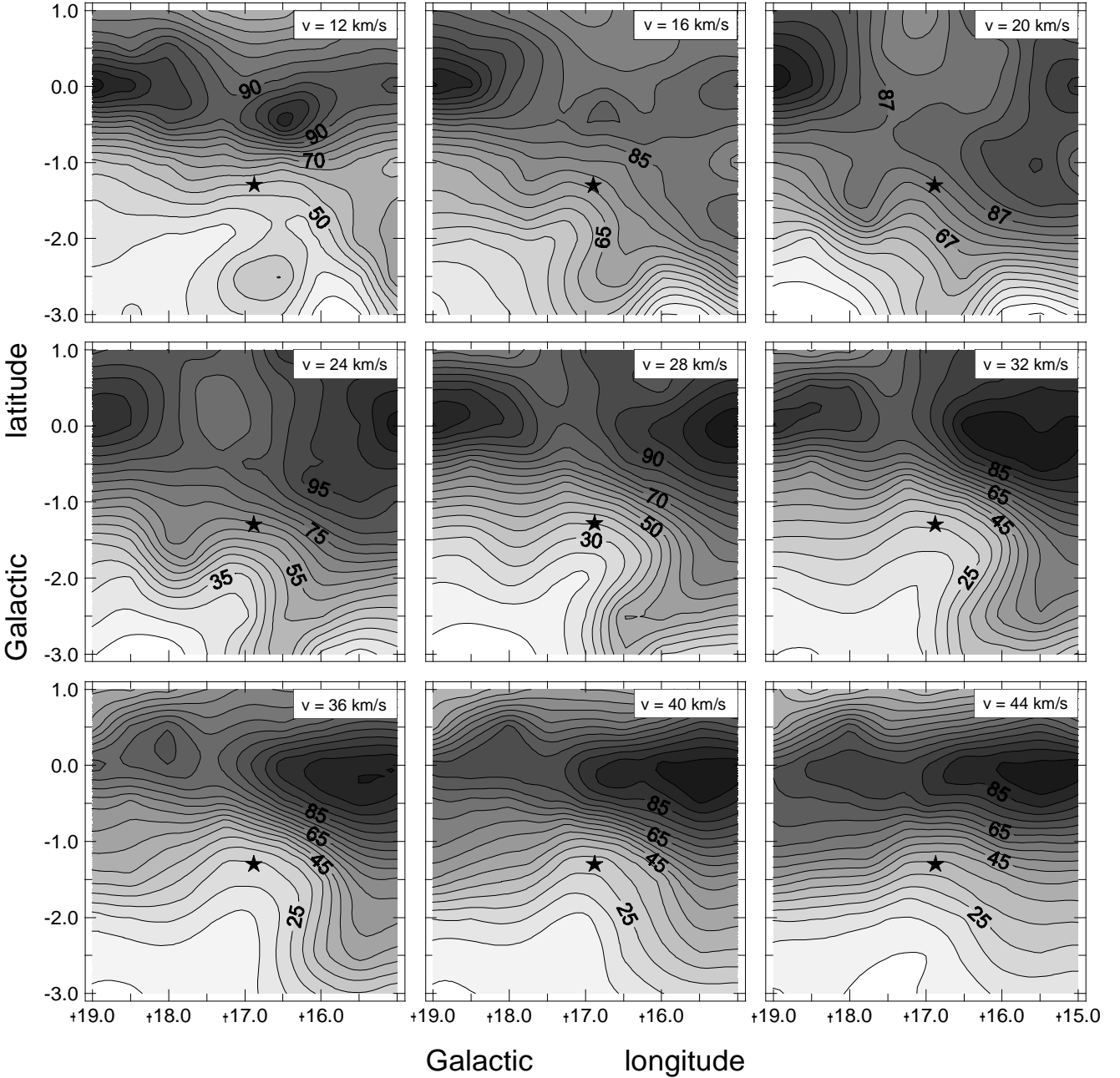}}
\caption{Neutral hydrogen channel maps for the velocity interval 10-46~km~s$^{-1}$, towards \object{LS~5039}. Each map covers 4~km~s$^{-1}$, the central velocity is given in each label. A local minimum and an open low density cavity can be clearly seen close to the HMXB. The cavity is open towards the direction opposite to the strongest mass concentrations near the galactic plane, as expected from a wind-blown structure.}
\label{nhi_chann}
\end{figure*}

Notwithstanding the absence of structures that can be clearly associated with
the SNR in the \ion{H}{i} distribution, the channel maps in the interval
10-46~km~s$^{-1}$ (see Fig.~\ref{nhi_chann}) reveal the existence of a large,
semi-open cavity which is very similar to bubbles blown out by early-type stars
in regions with steep density gradients (e.g., Benaglia \& Cappa
\cite{benaglia99}). This phenomenon has also been observed around another HMXB
(\object{HD~153919}) by Benaglia \& Cappa (\cite{benaglia99}). The ambient
material is thought to be swept up by the strong wind of the massive early-type
star in the binary system creating a local minimum around the star. Density
gradients towards the galactic plane in the original matter distribution can
result in a preferred escape direction for the wind, yielding open structures.
Even in some cases the star appears displaced from the actual minimum in the
\ion{H}{i} distribution.

%------------------------------------------------------------------------------
%------------------------------------------------------------------------------

The bubble detected around \object{LS~5039} is centered at a systemic velocity
of $v_{\rm sys}\simeq33$~km~s$^{-1}$, with an expansion velocity $v_{\rm
exp}\simeq12$~km~s$^{-1}$. The galactic rotation model by Fich et~al.
(\cite{fich89}) indicates that it is located at a distance of $\simeq3.2$~kpc,
quite consistent with the distance to the microquasar. The total mass removed
in order to create the cavity is $\sim1.5\,10^4M_{\sun}$, implying a kinetic
energy $\sim2\,10^{49}$ erg~s$^{-1}$, which can be considered as a rough
estimate of the energy deposited by the wind of the star into the ISM. Such a
value is in accordance with previous estimates for similar Of star wind blown
bubbles (e.g., Benaglia \& Cappa \cite{benaglia99}). With a radius of
$\sim60$~pc, the dynamical age of the bubble is $\sim3\,10^6$~yr, which should
be considered as an upper limit for its actual age since it ignores possible
contributions from the SN explosion and the wind of the progenitor of the
compact star.

We have also searched for other contributors to the cavity, among stars earlier
than B2 that were projected over the cavity, and whose distances were in
agreement with that of \object{LS~5039}. It seems unlikely that the Wolf Rayet
star WR~115 (WN6+OB?) could contribute in some way, because its distance is
2~kpc, and is located over the shell of the cavity ($l=16.98\degr$,
$b=-1.03\degr$). The other luminous stars in the field, \object{LS~5005},
\object{LS~5017}, \object{LS~5022}, \object{LS~5047}, \object{LS~5048} and
\object{HD~169673} can sum up at most 20\% of the energy required to create
such a minimum in the \ion{H}{i} distribution. The O((f)) star in
\object{LS~5039}, then, seems to be the main agent forming the bubble.

However, the actual space velocity indicates that the system has been close to
the cavity only during the last $\sim2\,10^5$~yr, which is much shorter than
the upper limit found for the dynamical age of the bubble ($\sim3\,10^6$~yr).
Nevertheless, the cavity could be originated by \object{LS~5039} if the pre-SN
system was close to the actual position, in such a way that the progenitor of
the compact object could have also contributed to the formation of the bubble.
In such a case, the X-ray binary would be only about $\sim2\,10^5$~yr old, and
we would expect to find the radio SNR originated in the SN event with a current
surface brightness of $\Sigma_{408}\sim 3\;10^{-22}$ W m$^{-2}$ Hz$^{-1}$
sr$^{-1}$ (Caswell \& Lerche \cite{caswell79}). Filtering techniques of the
diffuse galactic emission (e.g., Combi et al. \cite{combi98}) allow to detect
very low brightness SNRs with $\Sigma_{408}\sim10^{-22}$ W m$^{-2}$ Hz$^{-1}$
sr$^{-1}$, but nothing has been found in this region except for \object{SNR
G016.8$-$01.1}, whose surface brightness at 408 MHz would be $\Sigma_{408}\sim
6\;10^{-22}$ W m$^{-2}$ Hz$^{-1}$ sr$^{-1}$ for a typical spectral index
$\alpha=-0.5$. This level of flux is expected at this latitude for a younger
SNR, say with $t\sim 10^5$ yr, but in such a case the particle density of the
medium around \object{SNR G016.8$-$01.1} should be higher than the estimates
given in Sect. 5.2, with values of $n\sim50$ cm$^{-3}$, in order to confine the
SNR to its inferred size at $d\sim2.9$ kpc. These densities are 1 order of
magnitude higher than those derived from the \ion{H}{i} observations. However,
a clear rejection of \object{SNR G016.8$-$01.1} cannot be established on this
basis alone, due to the flux contamination from the \ion{H}{ii} region, which
could be responsible for errors as large as 100 \% in the flux density estimate
of the source. Consequently, a picture where the \ion{H}{i} bubble has been
created by both the stellar winds from the binary plus the SN explosion remains
as an open possibility.

\section{Discussion} \label{discussion}

There are two ways to accelerate a binary system by a supernova explosion:
ejection of material from the binary in a symmetric SN or an additional
velocity kick, produced by asymmetries in the SN itself. In Nelemans et~al.
(\cite{nelemans99}), the authors conclude that there is no need of additional
velocity kicks in systems like \object{Cygnus~X-1} or \object{GRO~J1655$-$40},
both containing black hole candidates as the compact objects, to explain their
observed large runaway velocities. On the other hand, from the study of
Be/X-ray binaries containing neutron stars, van den Heuvel et~al.
(\cite{vdheuvel00}) conclude that, in order to explain the observed large
eccentricities and low space velocities of these systems, a kick in an
asymmetric SN explosion scenario is needed. Moreover, the microquasar
\object{Circinus~X-1}, containing a neutron star, seems to have experienced a
highly asymmetric SN explosion (Tauris et~al. \cite{tauris99}). Unfortunately,
the nature of the compact object in \object{LS~5039} still remains unknown.
However, we can try to study it in the symmetric SN scenario.

Let $M_{\rm X}$ and $M_{\rm O}$ be the masses of the compact object and the
optical companion, respectively. Taking into account the stellar evolution
tracks by Schaerer et~al. (\cite{schaerer96}), we can assume that $M_{\rm
O}\simeq40\,M_{\sun}$. Considering that it must rotate slower than the breakup
speed, and using the mass function $f(m)=0.00103\pm0.00020~M_{\sun}$ by McSwain
et~al. (\cite{mcswain01}), an upper limit of $M_{\rm X}<9\,M_{\sun}$ is
obtained. The minimum mass for the compact object would be that of a neutron
star of the Chandrasekar mass, $M_{\rm X}>1.4\,M_{\sun}$. Hence, a total mass
of $\sim41$-$49\,M_{\sun}$ remains in the system.

Since the actual eccentricity of the system is $e=0.41\pm0.05$, and has a short
orbital period of $P=4.117\pm0.011$~days, both parameters are expected to
decrease with time due to tidal forces which act to re-circularize the orbit.
Hence, the eccentricity just after the SN explosion was $e_{\rm post-SN}>0.41$.
Using the equation $\Delta\,M=e_{\rm post-SN}\,(M_{\rm X}+M_{\rm O})$ and
$M_{\rm X}=1.4\,M_{\sun}$ we find that at least $17\,M_{\sun}$ where lost in
the SN explosion in the symmetric case. If we adopt $M_{\rm X}=9\,M_{\sun}$ the
minimum mass loss is $20\,M_{\sun}$. Moreover, in order to have the system
bound after the explosion, i.e., $e_{\rm post-SN}<1$, the mass loss during the
SN event must be less than the remaining mass in the system, i.e.,
$<49\,M_{\sun}$. Hence, the values for the reduced mass of the system, defined
as $\mu=1/(1+e_{\rm post-SN})$, range from 0.5 to 0.7. Using Eq.~(5) of
Nelemans et~al. (\cite{nelemans99}) with the actual values of $P$ and $e$ we
find that the re-circularized period will be $P_{\rm re-circ}\simeq3$~days.
Since the period of the initial, pre-SN, binary system is given by $P_{\rm
i}=\mu^2P_{\rm re-circ}$, the orbital period before the SN explosion was in the
range 0.8 to 1.5~days. Using Kepler's third law and assuming the latter value
to allow the maximum separation possible between the progenitor stars, initial
total masses between $60$-$100\,M_{\sun}$ lead to semimajor axis in the range
$22$-$26\,R_{\sun}$. This means that the two stars did not evolve separately,
but probably through a common envelope phase. Hence, a detailed evolution of
the system should be carried out in order to try to compute the initial mass of
the progenitor star of the compact object, the maximum age of the system before
the SN event, and the mass loss during the SN explosion itself. This kind of
study is beyond the scope of this paper.

Nevertheless, we can try to explain the measured $v_{\rm sys}$ in the context
of the symmetric SN explosion scenario using the following equation from
Nelemans et~al. (\cite{nelemans99}):
\begin{equation}
\left(\hspace{-1mm}{v_{\rm sys}\over{\rm km~s}^{-1}}
\hspace{-1mm}\right)\hspace{-0.5mm}=\hspace{-0.5mm}213\hspace{-1mm}\left(\hspace{-1mm}{\Delta\,M\over{M_{\sun}}}\hspace{-1mm}\right)\hspace{-1mm}\left(\hspace{-1mm}{M_{\rm
O}\over{M_{\sun}}}\hspace{-1mm}\right)\hspace{-1mm}\left(\hspace{-1mm}{P_{\rm
re-circ}\over{\rm
day}}\hspace{-1mm}\right)^{\hspace{-2mm}-{1\over3}}\hspace{-1mm}\left(\hspace{-1mm}{M_{\rm
X}\hspace{-1mm}+\hspace{-1mm}M_{\rm
O}\over{M_{\sun}}}\hspace{-1mm}\right)^{\hspace{-2mm}-{5\over3}}
\label{vsys}
\end{equation}
%\left(v_{\rm sys}\over{\rm km~s}^{-1}\right)=213\left(\Delta\,M\over{M_{\sun}}\right)\left(M_{\rm O}\over{M_{\sun}}\right)\left(P_{\rm re-circ}\over{\rm day}\right)^{-{1\over3}}\left(M_{\rm X}+M_{\rm O}\over{M_{\sun}}\right)^{-{5\over3}}
Assuming that $M_{\rm O}=40\,M_{\sun}$ and $P_{\rm re-circ}=3$~days we can find
two minimum values of $v_{\rm sys}$ depending on the adopted mass for the
compact object. For $M_{\rm X}=1.4\,M_{\sun}$, and hence
$\Delta\,M>17\,M_{\sun}$, we find $v_{\rm sys}\ga200$ km~s$^{-1}$. Similarly,
for $M_{\rm X}=9\,M_{\sun}$, and hence $\Delta\,M>20\,M_{\sun}$, we find
$v_{\rm sys}\ga180$ km~s$^{-1}$. Our measured recoil velocity is $v_{\rm
sys}\simeq150\pm20$ km~s$^{-1}$, which would be perfectly reached in the SN
explosion scenario, as we have just seen. In fact, the measured value is a
little bit lower than the computed ones, which could indicate that part of the
eccentricity of the system was reached thanks to a kick in an asymmetric SN
explosion. However, we must be cautious with the use of the value for $v_{\rm
sys}$ obtained after Eq.~\ref{vsys}, because some effects have been ignored, as
pointed out by Nelemans et~al. (\cite{nelemans99}), and also the errors in $e$
and $P_{\rm re-circ}$ could reduce the computed values. Hence, we can say that
both, the high eccentricity measured by McSwain et~al. (\cite{mcswain01}), and
the high space velocity reported here, are perfectly compatible, within errors,
with a symmetric SN explosion scenario. In other words, the actual eccentricity
of the system is naturally explained with the mass loss necessary to explain
the measured space (recoil) velocity of \object{LS~5039}.

Another common mechanism referred to produce runaway stars is based on
dynamical ejection from young open clusters. Numerical simulations (Kiseleva
et~al. \cite{kiseleva98}) show that it can explain some moderately
high-velocity stars observed in the Galactic disk, but only about 1\% of the
stars would be ejected with velocities higher than 30 km~s$^{-1}$. Hence, it
seems difficult to explain the high observed velocity of \object{LS~5039} as a
result of such a mechanism.

It is interesting to note that, contrary to other HMXB with low orbital
periods, \object{LS~5039} is still in the circularization process. Using the
actual value of $P$, a typical value of $R=10.3\,R_{\sun}$ for the radius of
the O6.5V star, and different formalisms (Claret et~al. \cite{claret95}; Claret
\& Cunha \cite{claret97}) we obtain typical circularization timescales of the
order of $(0.5\,$-$\,5)\,10^5$~yr. These values are compatible with the system
being still in the circularization process, since its formation just in the
galactic midplane would take $5\,10^5$~yr to bring it to the actual position
according the integration of its trajectory.

At this point we might ask what is the point of studying runaway microquasars.
In the case of \object{LS~5039}, if the SN explosion had taken place recently,
we could expect the system to survive for the following $\sim5$~Myr, because
the O6.5 star is still in the main sequence. Hence, according to the computed
trajectory, it would reach a height of $Z\simeq-600$~pc, which translates into
$b\simeq-12\degr$. On the other hand, \object{LS~5039} could be associated with
the high energy $\gamma$-ray source \object{3EG~J1824$-$1514}, as suggested by
Paredes et~al. (\cite{paredes00}). Hence, if this association is correct, we
could be able to detect $\gamma$-ray microquasars up to values of
$|b|>10\degr$, although all confirmed microquasars lie within $|b|<5\degr$. In
particular, runaway microquasars could be connected with some of the
unidentified faint, variable, and soft $\gamma$-ray EGRET sources above the
galactic plane, as suggested by Romero (\cite{romero01}) and Mirabel et~al.
(\cite{mirabel01}).

Hence, it would be interesting to study the proper motions and radial
velocities of as many microquasars as possible. Unfortunately, this kind of
information is not easy to obtain for a wide sample of systems. Proper motions
can be obtained either from optical data or either from radio data. However,
most optical counterparts of these sources are faint objects not present in old
astrometric catalogs. Hence, it is mandatory to acquire new positions. In this
context, it is better to do it in the radio domain because the uncertainties
are always smaller than those obtained in the optical. On the other hand, it is
necessary to carry out optical spectroscopy to obtain the radial velocity curve
of these sources and use the correct value for the radial velocity of the
system.

An alternative approach could be the search for signatures of bow shocks in the
microquasar vicinity. However, recent studies by Huthoff \& Kaper
(\cite{huthoff02}) of runaway OB stars, reveal that the success of this method
is highly dependant on the density of the ISM around the runaway object. In
particular, in only one object, namely \object{Vela~X-1}, a bow shock has been
detected. Finally, one could search for signatures of explosive events in the
ISM. Sensitive spectral line observations in the radio are probably the best
tool for this purpose, as done in \object{GRO~J1655$-$40} by Combi et~al.
(\cite{combi01}).

In any case, the study of runaway microquasars such as \object{LS~5039} is
likely to contribute significantly to different areas of modern high-energy
Astrophysics with independence of the observing technique.

\section{Conclusions} \label{conclusions}

After an in-depth study of the proper motions and surroundings of
\object{LS~5039} our main conclusions are:

\begin{enumerate}

\item Positions at optical and radio wavelengths have been used to compute
independent optical and radio proper motions, which are perfectly compatible.
Therefore, based only on astrometric data we are able to confirm that both, the
optical and the radio emission, originate in the same object.

\item From the combined optical and radio positions we have computed an
accurate proper motion for \object{LS~5039}. This, together with the new
estimate of $2.9\pm0.3$~kpc for the distance, allows us to compute a space
velocity of ($U=51$, $V=-71$, $W=-118$) km~s$^{-1}$ in its Regional Standard of
Rest (RSR). This results implies that \object{LS~5039} is a runaway microquasar
with $v_{\rm sys}\simeq150$ km~s$^{-1}$, escaping from its own RSR with a large
velocity component perpendicular to the galactic plane. This is probably the
result of the SN event that created the compact object in this binary system.

\item We have computed the past trajectory of \object{LS~5039}. Two OB
associations have been found close to its path in the plane of the sky.
However, they are too close to us to be related to the microquasar. On the
other hand, we have also found three SNRs near the path of \object{LS~5039}.
After discarding two of them based on distance arguments, we have focused our
attention on \object{SNR G016.8$-$01.1}. A study of this source could not
clearly confirm nor reject the association due to the large uncertainties in
the estimated radio flux density of the SNR. This fact perhaps justifies
future, high sensitivity searches of low-brightness remnants in this region.

\item We have found a semi-open \ion{H}{i} cavity close to the \object{LS~5039}
position. Although the O((f)) star in this microquasar seems to be the main
agent forming the bubble, a contribution from the progenitor of the compact
object cannot be ruled out.

\item Finally, we are able to explain both, the high space velocity and the
high eccentricity observed, in a symmetric SN explosion scenario with a mass
loss of $\Delta\,M\sim17\,M_{\sun}$.

\end{enumerate}

\begin{acknowledgements}

We are grateful to S.~E. Urban for kindly providing information from the
AC~2000.2 catalog, and for useful discussion on the Tycho-2 proper motion.
We are also grateful to D.~G. Monet and N. Zacharias for kindly providing
information on USNO-A2.0 and UCAC1 catalogs, respectively.
M.~R. acknowledges useful comments and discussions with I. Ribas, F. Figueras, D. Fern\'andez, J. Colom\'e and E. Masana, members of Departament d'Astronomia i Meteorologia, at Universitat de Barcelona.
%We acknowledge detailed and useful comments from an anonymous referee.
This research has made use of the SIMBAD database, operated at CDS, Strasbourg,
France.
The authors acknowledge the data analysis facilities provided by the Starlink
Project which is run by CCLRC on behalf of PPARC.
The Guide Star Catalog was produced at the Space Telescope Science Institute
under U.S. Government grant.
M.~R. is supported by a fellowship from CIRIT (Generalitat de Catalunya, ref.
1999~FI~00199).
J.~M.~P., J.~M. and M.~R. acknowledge partial support by DGI of the Ministerio de Ciencia y Tecnolog\'{\i}a (Spain) under grant AYA2001-3092.
G.~E.~R. is supported by the research grants PICT 03-04881 (ANPCT) and PIP
0438/98 (CONICET), as well as by Fundaci\'on Antorchas. 
He is very grateful to staff of the Max Planck Institut f\"ur Kernphysik at
Heidelberg, where part of his research for this project was carried out.
J.~M has also been aided in this work by an Henri Chr\'etien International
Research Grant administered by the American Astronomical Society.

\end{acknowledgements}


\begin{thebibliography}{}

\bibitem[2001]{ankay01}
Ankay, A., Kaper, L., de Bruijne, et~al.
2001, A\&A, 370, 170

\bibitem[1999]{benaglia99}
Benaglia, P.~\& Cappa, C.~E.
1999, A\&A, 346, 979

\bibitem[1979]{caswell79}
Caswell, J.~L.~\& Lerche, I.
1979, MNRAS, 187, 201

\bibitem[2001]{castro01}
Castro-Tirado, A.~J., Greiner, J., \& Paredes, J.~M.
2001, Proc. of the Third Microquasar Workshop 'Galactic Relativistic Jet Sources', ed. A.~J. Castro-Tirado, J. Greiner, and J.~M. Paredes, Kluwer Academic Publishers

\bibitem[1997]{claret97}
Claret, A.~\& Cunha, N.~C.~S.
1997, A\&A, 318, 187

\bibitem[1995]{claret95}
Claret, A., Gim\'enez, A., \& Cunha, N.~C.~S.
1995, A\&A, 299, 724

\bibitem[2001]{clark01}
Clark, J.~S., Reig, P., Goodwin, S.~P., et~al.
2001, A\&A, 376, 476

\bibitem[1998]{combi98}
Combi, J.~A., Romero, G.~E., \& Benaglia, P.
1998, A\&A, 333, L91

\bibitem[2001]{combi01}
Combi, J.~A., Romero, G.~E., Benaglia, P., \& Mirabel, I.~F.
2001, A\&A, 370, L5

\bibitem[1998]{condon98}
Condon, J.~J., Cotton, W.~D., Greisen, E.~W., et~al.
1998, AJ, 115, 1693

\bibitem[1995]{dauphole95}
Dauphole, B.~\& Colin, J.
1995, A\&A, 300, 117

\bibitem[1976]{deutschman76}
Deutschman, W.~A., Davis, R.~J., \& Schild, R.~E.
1976, ApJS, 30, 97

\bibitem[1991]{drilling91}
Drilling, J.~S.
1991, ApJS, 76, 1033

\bibitem[1999]{dubner99}
Dubner, G., Giacani, E., Reynoso, E., et~al.
1999, AJ, 118, 930

\bibitem[1989]{fich89}
Fich, M., Blitz, L., \& Stark, A.~A.
1989, ApJ, 342, 272

\bibitem[2000]{forbes00}
Forbes, D.
2000, AJ, 120, 2594

\bibitem[2000]{green00}
Green, D.~A.
2000, A Catalog of Galactic Supernova Remnants, Mullard Radio Astronomy
Observatory, Cambridge, England, UK (available on the World Wide Web at
http://www.mrao.cam.ac.uk/surveys/snrs/)

\bibitem[2001]{hannikainen01}
Hannikainen, D., Campbell-Wilson, D., Hunstead, R. et~al.
2001, in Proc. of the Third Microquasar Workshop 'Galactic Relativistic Jet Sources', ed. A.~J. Castro-Tirado, J. Greiner, and J.~M. Paredes, Kluwer Academic Publishers, Ap\&SS, 276, 45

\bibitem[2000]{hog00}
H{\o}g, E., Fabricius, C., Makarov, V.~V., et~al.
2000, A\&A, 357, 367

\bibitem[2002]{huthoff02}
Huthoff, F.~\& Kaper, L.
2002, A\&A, in press

\bibitem[1966]{johnson66}
Johnson, H.~L.
1966, ARA\&A, 4, 193

\bibitem[1999]{kaper99}
Kaper, L., Camer\'on, A., \& Barziv, O.
1999, in Proc. of the IAU Symp.~193: Wolf-Rayet phenomena in massive stars and starburst galaxies, ed. K.~A. van der Hucht, G. Koenigsberger, and P.~R.~J. Eenens, ASP Conference Series, 193, p. 316

\bibitem[1998]{kiseleva98}
Kiseleva, L.~G., Colin, J., Dauphole, B., \& Eggleton, P.
1998, MNRAS, 301, 759

\bibitem[1992]{lahulla92}
Lahulla, J.~F.~\& Hilton, J.
1992, A\&AS, 94, 265

\bibitem[1990]{lasker90}
Lasker, B.~M., Sturch, C.~R., McLean, B.~J., et~al.
1990, AJ, 99, 2019

\bibitem[2001]{lejeune01}
Lejeune, T.~\& Schaerer, D.
2001, A\&A, 366, 538

\bibitem[2000]{liu00}
Liu, Q.~Z., van Paradijs, J., \& van den Heuvel, E.~P.~J.
2000, A\&AS, 147, 25

\bibitem[2001]{liu01}
Liu, Q.~Z., van Paradijs, J., \& van den Heuvel, E.~P.~J.
2001, A\&A, 368, 1021

\bibitem[1998]{marti98}
Mart\'{\i}, J., Paredes, J.~M., \& Rib\'o, M.
1998, A\&A, 338, L71

\bibitem[2001]{mcswain01}
McSwain, M.~V., Gies, D.~R., Riddle, R.~L., Wang, Z., \& Wingert, D.~W.
2001, ApJ, 558, L43

\bibitem[1997]{melnik97}
Melnik, A.~M.~\& Efremov, Y.~N.
1997, VizieR Online Data Catalog, 902, 110013

\bibitem[1999]{mirabel99}
Mirabel, I.~F.~\& Rodr\'{\i}guez, L.~F.
1999, ARA\&A, 37, 409

\bibitem[2001]{mirabel01}
Mirabel, I.~F., Dhawan, V., Mignani, R.~P., Rodrigues, I., \& Guglielmetti, F.
2001, Nature, 413, 139

\bibitem[1999]{monet99}
Monet, D.~G., Bird, A., Canzian, B., et~al.
1999, USNO-A2.0 CD-ROM (U.S. Naval Observatory, Washington DC)

\bibitem[2001]{morrison01}
Morrison, J.~E., R{\"o}ser, S., McLean, B., Bucciarelli, B., \& Lasker, B.
2001, AJ, 121, 1752

\bibitem[1997]{motch97}
Motch, C., Haberl, F., Dennerl, K., Pakull, M., \& Janot-Pacheco, E.
1997, A\&A, 323, 853

\bibitem[1999]{nelemans99}
Nelemans, G., Tauris, T.~M., \& van den Heuvel, E.~P.~J.
1999, A\&A, 352, L87

\bibitem[2000]{ochsenbein00}
Ochsenbein, F., Bauer, P., \& Marcout, J.
2000, A\&AS, 143, 23

\bibitem[2000]{paredes00}
Paredes, J.~M., Mart\'{\i}, J., Rib\'o, M., \& Massi, M.
2000, Science, 288, 2340

\bibitem[2000]{reed00}
Reed, B.~C.
2000, AJ, 120, 314

\bibitem[1986]{reich86}
Reich, W., Fuerst, E., Reich, P., Sofue, Y., \& Handa, T.
1986, A\&A, 155, 185

\bibitem[2000]{reynoso00}
Reynoso, E.~M.~\& Mangum, J.~G.
2000, ApJ, 545, 874

\bibitem[1999]{ribo99}
Rib\'o, M., Reig, P., Mart\'{\i}, J., \& Paredes, J.~M.
1999, A\&A, 347, 518

\bibitem[2002]{ribo02}
Rib\'o, M., Paredes, J.~M., Mart\'{\i}, J., \& Massi, M.
2002, in preparation

\bibitem[1960]{rodgers60}
Rodgers, A.~W., Campbell, C.~T., \& Whiteoak, J.~B.
1960, MNRAS, 121, 103

\bibitem[2001]{romero01}
Romero, G.~E.
2001, in Proc. of The Nature of Unidentified Galactic High-energy Gamma-ray Sources, ed. A. Carrami\~nana, O. Reimer, and D.~J. Thompson, ASSL Series of Kluwer Academic Publishers 51, p. 65

\bibitem[1996]{schaerer96}
Schaerer, D., de Koter, A., Schmutz, W., \& Maeder, A.
1996, A\&A, 312, 475

\bibitem[1982]{schmidt82}
Schmidt-Kaler, Th.
1982, Landolt-B\"ornstein: Numerical Data and Functional Relationships in
Science and Technology, New Series "Group 6 Astronomy and Astrophysics", Volume 2

\bibitem[1959]{sedov59}
Sedov, L.~I.
1959, Similarity and Dimentional Methods in Mechanics, John Wiley and Sons,
New York

\bibitem[1999]{shahbaz99}
Shahbaz, T., van der Hooft, F., Casares, J., Charles, P.~A., \& van Paradijs, J.
1999, MNRAS, 306, 89

\bibitem[1968]{shklovskii68}
Shklovskii, I.~S.
1968, Supernovae, John Wiley and Sons, New York

\bibitem[1998]{spitzer98}
Spitzer, L.
1998, Physical Processes in the Interstellar Medium, John Wiley and Sons,
New York

\bibitem[2001]{stirling01}
Stirling, A.~M., Spencer, R.~E., de la Force, C.~J., et~al.
2001, MNRAS, 327, 1273

\bibitem[1994]{tasker94}
Tasker, N.~J., Condon, J.~J., Wright, A.~E., \& Griffith, M.~R.
1994, AJ, 107, 2115

\bibitem[1999]{tauris99}
Tauris, T.~M., Fender, R.~P., van den Heuvel, E.~P.~J., Johnston, H.~M., \& Wu, K.
1999, MNRAS, 310, 1165

\bibitem[2000]{torra00}
Torra, J., Fern{\'a}ndez, D., \& Figueras, F.
2000, A\&A, 359, 82

\bibitem[1998]{urban98}
Urban, S.~E., Corbin, T.~E., Wycoff, G.~L., et~al.
1998, AJ, 115, 1212

\bibitem[1996]{vacca96}
Vacca, W.~D., Garmany, C.~D., \& Shull, J.~M.
1996, ApJ, 460, 914

\bibitem[1978]{vdheuvel78}
van den Heuvel, E.~P.~J., Portegies Zwart, S.~F., Bhattacharya, D., \& Kaper, L.
2000, A\&A, 364, 563

\bibitem[2000]{vdheuvel00}
van den Heuvel, E.~P.~J.
1978, in Physics and Astrophysics of Neutron Stars and Black Holes, ed. R. Giacconi and R. Ruffini (Amsterdam: North Holland), 828

\bibitem[1999]{zacharias99a}
Zacharias, N.~\& Zacharias, M.~I.
1999, AJ, 118, 2503

\bibitem[1999]{zacharias99b}
Zacharias, N., Zacharias, M.~I., \& de Vegt, C.
1999, AJ, 117, 2895

\bibitem[2000]{zacharias00}
Zacharias, N., Urban, S.~E., Zacharias, M.~I., et~al.
2000, AJ, 120, 2131

\end{thebibliography}
\end{document}